\documentclass{iopart}
\usepackage{epsfig}

\newcommand{\beq}{\begin{equation}}
\newcommand{\eeq}{\end{equation}}

\begin{document}

\title{Vortices in a Bose-Einstein condensate confined by an optical lattice}
\author{P.G.\ Kevrekidis$\dag$, R.\ Carretero-Gonz\'{a}lez$\ddag$, 
G.\ Theocharis$\S$, D.J.\ Frantzeskakis$\S$, and B.A.\ Malomed$\parallel$}

\begin{abstract}
We investigate the dynamics of vortices in repulsive Bose-Einstein condensates
in the presence of an optical lattice (OL) and a parabolic magnetic trap.
The dynamics is sensitive to the phase of the OL potential relative to the
magnetic trap, and depends less on the OL strength. For the cosinusoidal OL
potential, a local minimum is generated at the trap's center, creating a
stable equilibrium for the vortex, while in the case of the sinusoidal
potential, the vortex is expelled from the center, demonstrating spiral
motion. Cases where the vortex is created far from the trap's center are also
studied, revealing slow outward-spiraling drift. 
Numerical results are explained in an
analytical form by means of a variational approximation. Finally, motivated
by a discrete model (which is tantamount to the case of the strong OL
lattice), we present a novel type of vortex consisting of two pairs of
anti-phase solitons.
\end{abstract}

\address{$\dag$ Department of Mathematics and Statistics,
University of Massachusetts, Amherst MA 01003-4515, USA}

\address{$\ddag$ Nonlinear Dynamical Systems Group\footnote[5]{http://nlds.sdsu.edu/},
Department of Mathematics and Statistics, San Diego State University, San Diego CA, 92182-7720} 

\address{$\S$ Department of Physics, University of Athens, Panepistimiopolis, 
Zografos, Athens 15784, Greece}

\address{$\parallel$ Department of Interdisciplinary Studies, Faculty of
Engineering, Tel Aviv University, Tel Aviv 69978, Israel}


\bigskip

{J.\ Phys.\ B: At.\ Mol.\ Phys.\ {\bf 36} (2003) 3467--3476.}

\maketitle

%
%
%
%
%



The experimental realization and theoretical studies of Bose-Einstein
condensates (BECs) \cite{review} have led to an explosion of interest in the
field of atomic matter waves and their nonlinear excitations, including dark 
\cite{dark} and bright \cite{bright} solitons. More recently,
two-dimensional (2D) excitations, such as vortices \cite{vortex} and vortex
lattices \cite{vl}, were considered and realized experimentally. Other
nonlinear states, such as e.g., Faraday waves \cite{stal}, ring dark
solitons and vortex necklaces \cite{theo}, stable solitons and localized
vortices in attractive BECs trapped in a 2D optical lattice (OL) \cite
{Salerno}, and even stable solitons supported by an OL in a \emph{repulsive}
BEC \cite{Baizakov} were also predicted.

Vortices, in particular, are worth studying not only due to their
significance as a fundamental type of coherent nonlinear excitations, but
also because they play a dominant role in the breakdown of superflow in Bose
fluids \cite{jac11,jac12}. The theoretical description of vortices in BECs
can be carried out in a much more efficient way than in liquid He \cite{jac1}
due to the weakness of interactions in the former case (which, in addition,
is tunable \cite{inouye}). These advantages explain a large volume of work
regarding the behavior of vortices in BECs, which has been recently
summarized in a review \cite{fetter}.

The subject of the present paper is the dynamics of \ vortices under the
action of the OL; to the best of our knowledge, this problem is considered in
this work for the first time. An OL potential is generated as an
interference pattern by laser beams illuminating the condensate, in
particular in 1D and 2D cases \cite{
greiner,tromb,konot,tromb2,catal2}.
In the 2D case, the OL potential assumes the form (in dimensionless units) 
\begin{equation}
V_{\mathrm{OL}}(x,y)=V_{0}\left[ \cos ^{2}(kx+\phi )+\cos ^{2}(ky+\phi )
\right] ,  \label{veq1}
\end{equation}
where $V_{0}$ is the strength of the OL, which is measured in units of the 
recoil energy $E_{r}=h^2/2m\lambda_{\mathrm{laser}}^{2}$ 
(i.e., the kinetic energy gained by 
an atom when it absorbs a photon from the OL), where $\lambda_{\mathrm{laser}}$
is the laser wavelength, $h$ is Planck's constant, $m$ is the atomic 
mass, $k$ is the wavenumber of the OL, and $\phi $ is a
phase-detuning factor [the obvious possibility to remove $\phi $ by means of
the diagonal shift, $\left( x,y\right) \rightarrow \left( x-\phi /k,y-\phi
/k\right) $, is ruled out by the presence of the magnetic-trap potential,
see below].
The wavenumber $k$ of the OL can be experimentally controlled by varying the
angle between the counter-propagating lasers producing the
interference pattern \cite{Morsch-Arimondo}.

As is well known, the effective 2D GP equation applies to
situations when the condensate has a nearly planar (``pancake'') shape, see,
e.g., \cite{Beersheva} and references therein. Accordingly, vortex
states considered below are not subject to 3D instabilities (corrugation of
the vortex axis \cite{fetter}) as the transverse dimension is effectively
suppressed.
We base the analysis on the Gross-Pitaevskii (GP) equation, written in
harmonic-oscillator units \cite{rupr}, 
\begin{equation}
iu_{t}=-\Delta u+|u|^{2}u+V(x,y)\,u,  \label{veq2}
\end{equation}
where $u=u(x,y,t)$ is the 2D wave function, and the external potential is 
\begin{equation}
V(x,y)=\frac{1}{2}\Omega^{2} (x^{2}+y^{2})+V_{\mathrm{OL}}(x,y),  \label{veq3}
\end{equation}
which includes the isotropic magnetic trap \cite{review} and OL potential 
(\ref{veq1}). Notice that the dimensionless parameter 
$\Omega \equiv \omega_{r}/\omega_{z}$ in Eq.\ (\ref{veq3}), 
where $\omega_{r,z}$ are the confinement frequencies in the radial and 
axial directions respectively, is assumed to be $\Omega  \ll 1$. 
In most studies of vortices, angular momentum imparted by stirring 
of the condensate is assumed (see, e.g., \cite{perez,jackson} 
and references therein), as typically this is the way in which vortices
are generated in experimental settings \cite{vortex}. 
Here we assume that a vortex has been generated, but
then the stirring ceases. Alternatively, the vortex may be created by means
of the phase-engineering technique \cite{Williams:99}. The positive sign in
front of the nonlinear term in Eq.\ (\ref{veq2}) implies that we consider a
repulsive condensate.

We first present numerical findings for the case of a ``regular'' vortex;
subsequently, the observed dynamics is explained by means of a variational
approximation. We then proceed to examine a new type of a vortex, which
is a robust bound state of two pairs of $\pi$-out-of-phase pulses, as
suggested by results known for 2D dynamical lattices. In most cases, we fix
the strength of the magnetic-trap potential to be $\Omega ^{2}=0.002$. 
This value is relevant to a $^{87}$Rb condensate of radius $25\mu$m, 
containing $4.6\times 10^{4}$ atoms in a highly anisotropic trap with 
$\omega_{r}=2\pi \times 7.5$ Hz and $\omega_{z}=2\pi \times 115$ Hz. 
According to these values of the physical parameters, in the following 
results, which will be presented in normalized time and space units, the 
corresponding units are $1.38$ ms and $1\mu$m respectively. Additionally, 
as far as the OL parameters are concerned, the typical values $V_{0}=0.5$ 
and $k=1$ used in most cases, correspond to $0.5 E_{r}$ and to an OL
wavelength of $6.3\mu$m 

In simulations, the initial vortex configuration is taken, in polar
coordinates $\rho $ and $\theta $, as 
\begin{equation}
u=\rho (r)\,\exp \left( i\theta \right) \,u_{\mathrm{TF}},  \label{veq5}
\end{equation}
where $\rho (r)=r^{2}(0.34+0.07r^{2})/(1+0.41r^{2}+0.07r^{4})$ is\ the
radial Pad{\'{e}} interpolation for the vortex solution to the GP equation
without external potential, and $u_{\mathrm{TF}}=\sqrt{\max \left\{ 0,\mu
-\left( \Omega ^{2}/2\right) (x^{2}+y^{2})\right\} }$ is the Thomas-Fermi
(TF) wave function for the magnetic trap \cite{review}, with a chemical
potential $\mu $. The vortex is placed at the center of the magnetic trap
(unless otherwise indicated). Simulations were performed by means of a
finite-difference discretization in a box of the size $60\times 60$
($60\mu$m$\times 60\mu$m), using 
$200\times 200$ points (i.e., the spatial stepsizes are $dx=dy=0.3$). Time
integration was performed by means of the fourth-order Runge-Kutta scheme
with the time step $dt=0.0025$ ($dt=3.45\mu$s). It should be remarked
that in the initial stages of the time evolution, the initial condition
of Eq. (\ref{veq5}) ``adjusts'' itself to the presence of the OL, by 
shedding small amplitude radiation waves. These are dissipated by the 
an absorbing region close to the boundary implemented in the numerical
simulation.

\begin{figure}[tbp]
\epsfxsize=9cm \centerline{\epsffile{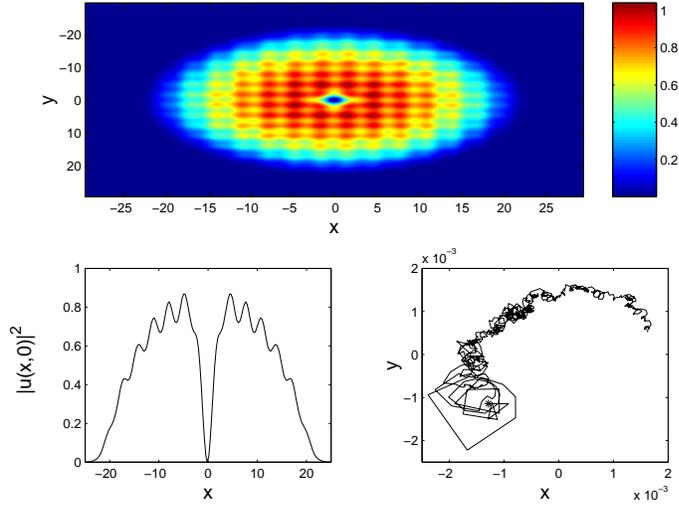}}
\caption{The vortex on a BEC of diameter $\approx 50\mu$m 
in the cosinusoidal [$\protect\phi =0$ in Eq.\ (\ref {veq1})] OL potential 
with $V_{0}=0.5$ and $k=1$. The top panel shows the contour plot of the 
density at $t=100$ ($138$ ms). The bottom left panel is a cut of
the same density profile along $x=0$, while the bottom right one shows the
motion of the vortex center for $0\leq t\leq 100$, the initial position
being marked by a star. Notice the scale ($10^{-3}$, or $1$ nm in physical units) 
of very weak jiggling
of the vortex, which thus stays practically immobile at the origin. }
\label{vfig1}
\end{figure}

The first case that was examined is the one with $\phi =0$ in Eq.\ (\ref
{veq1}). In this case, the vortex is extremely robust, staying at the center
during a few hundred time units of Eq.\ (\ref{veq2}), which correspond to a 
few hundred of milliseconds in physical units. As can be seen in 
Fig.\ \ref{vfig1}, where the
vortex and its motion are shown for $V_{0}=0.5$ and $k=1$, the displacement
of the vortex center remains, indefinitely long, as small as $\sim 10^{-3}$.
The center of the vortex was located by fitting, close to the local minimum,
the density $|u|^{2}$ to the expression $|u|^{2}=Ax^{2}+By^{2}+Cx+Dy+E$,
which yields the position of the center at $(-C/2A,-B/2D)$).

\begin{figure}[tbp]
\epsfxsize=9cm \centerline{\epsffile{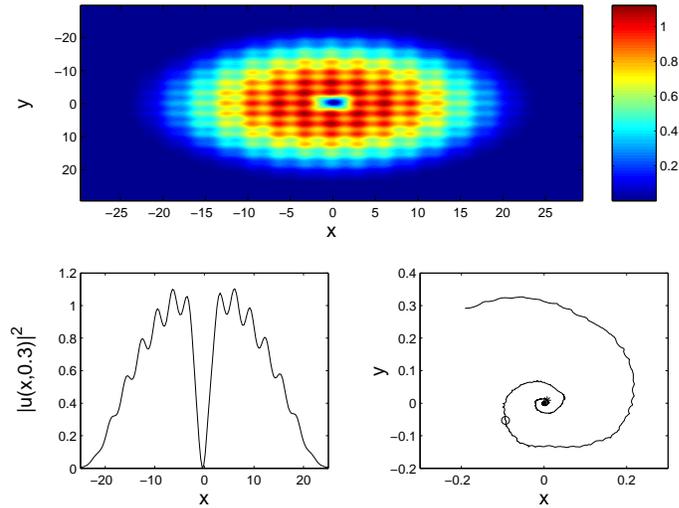}}
\caption{The same as in Fig.\ \ref{vfig1}, but now the potential is
sinusoidal, $\protect\phi =\protect\pi/2$. To clearly demonstrate the
dynamics, the top and bottom left panels show the density profiles at a
larger time than in Fig.\ \ref{vfig1}, $t=250$ ($346$ ms). The bottom right panel
depicts the motion of the vortex center for $0\leq t\leq 250$ [positions of
the vortex center at $t=100$ ($138$ ms) and $t=200$ ($276$ ms), 
respectively, are indicated by the star and circle].}
\label{vfig2}
\end{figure}

In Fig.\ \ref{vfig2}, we display the evolution which starts from the same
initial configuration, but with $\phi =\pi /2$ in Eq.\ (\ref{veq1}). A
distinctly different behavior is observed in this case, which is shown for
longer times, up to $t=250$ ($346$ ms), to verify that it is a real feature. In
particular, it is clearly observed in this case that the vortex \textit{
spirals out} from the center of the magnetic trap. Note that fine details of
the motion indicate meandering motion, similar to that exhibited by spiral
waves in excitable media \cite{bjorn}, even though in the present case the
meandering occurs on a smaller spatial scale, hence much finer resolution
would be needed to clarify it. However, the outward-spiraling motion of the
vortex center is clearly seen in Fig.\ \ref{vfig2}. Considerably longer runs
[for times on the order of a few thousand time units of Eq.\ (\ref{veq2}),
which correspond to seconds in physical units]
show that the vortex spirals all the way to the edges of the TF cloud where
it eventually decays (see also comments below).

\begin{figure}[tbp]
\epsfxsize=9cm \centerline{\epsffile{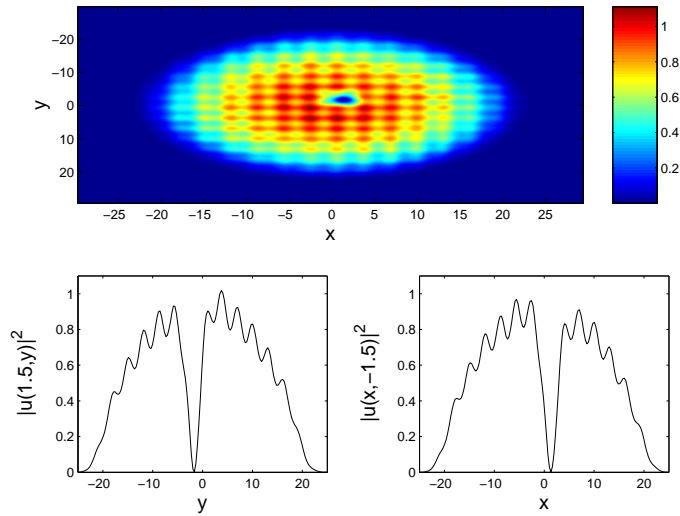}}
\caption{The case of $V_{0}=0.5$ and $\protect\phi =\protect\pi/4$ in Eq. 
(\ref{veq1}). 
The vortex
moves rapidly away from the center. The snapshots show the density profile
(top panel) and its horizontal and vertical cuts (along $x=1.5$ and 
$y=-1.5$, respectively) of the field at $t=100$ ($138$ ms).}
\label{vfig3}
\end{figure}

In Fig.\ \ref{vfig3}, we examine the case with $\phi =\pi /4$ in Eq.\ (\ref
{veq1}). In this case also, the vortex moves outward. In fact, it is
observed that it moves away from the trap center much faster, as at $t=100$ ($138$ ms) 
it is already located near $(x,y)=(1.5,-1.5)$ (cf.\ Fig.\ \ref{vfig2}),
corresponding to a couple of $\mu$m from the trap center.

Qualitatively similar conclusions, but at different time scales, have been
obtained for different values of the OL strength $V_{0}$. For instance, in
the potential with $\phi =\pi /2$, $V_{0}=1.5$ (and $k=1$), the vortex
spirals out again (similar to the case with $V_{0}=0.5$ displayed above),
the time necessary for it to reach the distance of $0.1$ from the origin
being three times as large as in the case of $V_{0}=0.5$ (details are not
shown here). This can be explained by the fact that, if $V_{0}$ is larger,
the vortex has to move through a ``rougher'' energy landscape, hence it
becomes more difficult for it to ``find its way out''.

These findings can be qualitatively 
understood in terms of an effective potential which
governs the motion of the vortex; in particular, the potential has a minimum
and maximum at the center of the magnetic trap in the cases of $\phi =0$ and 
$\phi =\pi /2$, respectively. To present this explanation in a
mathematically tractable
form, we resort to a variational approximation. 
As in Ref.\ \cite{tempere}, we use the following 
\textit{ansatz} for the vortex, inspired by its linear analog namely the 
first excited state of the quantum harmonic oscillator: 
%
%
\begin{equation}
u(x,y,t)=B(t)\,r_{0}(t)\exp \left[ -r_{0}^{2}(t)/b(t)\right] \,\,
e^{i\varphi_{0}(t)},  \label{veq6}
\end{equation}
where $r_{0}^{2}(t)\equiv \left( x-x{_{0}(}t)\right) ^{2}+\left(
y-y_{0}(t)\right) {^{2}}$, and $\varphi_{0}(t)\equiv \tan
^{-1}[(y-y_{0}(t))/(x-x_{0}(t))]$.
Note that the ansatz carries a vortex-like structure centered at 
$(x_{0},y_{0})$, and it qualitatively emulates the initial waveform (\ref
{veq5}) which was adopted in the numerical simulations. The stronger the
nonlinearity, the less accurate the approximation offered by Eq.\ 
(\ref{veq6}) is; however, as the ansatz (\ref{veq6}) was found 
to always provide a
qualitatively correct description of the phenomenology, we use it to
represent the vortex.

Similarly to the calculations reported in Refs.\ \cite{ricardo1,ricardo2},
one can easily deduce (additionally using the norm conservation) that the
evolution of $B(t)$ and $b(t)$ is, to the leading order, negligible,
therefore we set them to constant values, $B\approx B(t)$ and 
$b\approx b(t)$. Then, the substitution of the ansatz (\ref{veq6}) 
into the Lagrangian of
the GP equation (\ref{veq2}) with the potential (\ref{veq3}) leads, up to
constant factors, to the effective Lagrangian 
\begin{equation}
L=\frac{1}{2}\left( \dot{x}_{0}^{2}+\dot{y}_{0}^{2}\right) 
-V_{\mathrm{eff}}(x_{0},y_{0}),  \label{veq7}
\end{equation}
where the net effect of the OL and magnetic trap is combined into an
effective potential, 
\begin{equation}
V_{\mathrm{eff}}(x,y)=Q(\phi )\left[ \cos (2kx)+\cos (2ky)\right] +
\frac{1}{4}\Omega^{2} \left( x^{2}+y^{2}\right),  
\label{Veff}
\end{equation}
where $Q(\phi )$ is given by a rather cumbersome expression; we will
actually need the values 
%
$$
\begin{array}{lll}
\displaystyle
Q(0)&\!\!\!\!=\!\!\!\!&\displaystyle\frac{V_0}{8}\, (bk^{2}-2)\, 
       \exp (-bk^{2}/{2}),\\[2.0ex]
\displaystyle Q\left(\frac{\pi}{2}\right)&\!\!\!\!=\!\!\!\!&-Q(0), ~~\mbox{and}\\[2.0ex]
\displaystyle Q\left(\frac{\pi}{4}\right)&\!\!\!\!=\!\!\!\!&0.
\end{array}
$$

\begin{figure}[tbp]
\epsfxsize=9cm \centerline{\epsffile{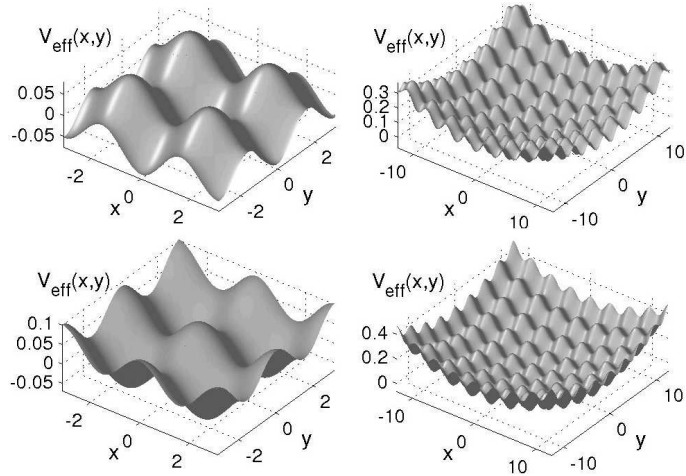}} 
\caption{The effective potential $V_{\mathrm{eff}}(x,y)$, obtained in the
variational approximation [cf.\ Eq.\ (\ref{Veff})], responsible for the
motion of the vortex subject to the 2D optical lattice and magnetic trap.
Top panel: the cosinusoidal optical lattice ($\protect\phi =0$) produces a
minimum in the effective potential at the origin. Bottom: the sinusoidal
optical potential ($\protect\phi =\protect\pi /2$) gives rise to a maximum
of the potential at the origin. The right panels display more clearly the
combined effect of the optical lattice and magnetic trap (here we use 
$\Omega ^{2}=0.005$, instead of $\Omega ^{2}=0.002$, which was used in the
computations). The parameters are $b=1$, $k=1$, $V_{0}=0.5$. }
\label{vfig4}
\end{figure}

Equations (\ref{veq7}) and (\ref{Veff}) suggest that the motion of the
coordinates $x_{0}$ and $y_{0}$ of the vortex core resembles the motion of
two uncoupled oscillators. Actually, this is a straightforward
generalization of a result that can be obtained in the corresponding 1D
case: in that case, the counterpart of the vortex is a dark soliton, whose
equation of motion in the presence of the potential (\ref{veq3}) can be
derived by means of the adiabatic perturbation theory for dark solitons in
BECs; see e.g., \cite{frantz} and references therein. 
This approach, in the presence of the one-dimensional magnetic trap and OL, 
yields an effective potential 

\[
V_{\mathrm{eff}}^{\mathrm{(1D)}}(x)=
\frac{1}{4}V_{0}\left[1-\frac{1}{6} 
\left(\frac{\pi^2}{3}-2\right)k^{2} \right] \cos\left(2k x\right)+
\frac{1}{4}\Omega^{2} x^{2}.
\]
In fact, Eq.\ (\ref{Veff}) is a generalization of this expression. The
effective potential from Eq.\ (\ref{Veff}) is depicted in Fig.\ \ref{vfig4}
as a function of $x$ and $y$ for the cases $\phi =0$ and $\phi =\pi /2$,
where we set $b=1$ [this value was chosen, comparing the size of the
numerically obtained vortex with that implied by the ansatz (\ref{veq6})].

In our experiments $k=1$ was used, and thus, in
the case $\phi =0$, a vortex positioned at the center of the magnetic
trap is at a local minimum of the potential (\ref{Veff}), therefore the
vortex stays in this position. If slightly perturbed around this stable
fixed point, the vortex will perform small amplitude oscillations.
On the contrary, in the case of $\phi =\pi /2$, the same point 
$(x_{0},y_{0})=(0,0)$ is unstable (as a saddle), and as a result the vortex
moves away from it. These conclusions are in full accord with the 
numerical experiments, see Figs. \ref{vfig1} and \ref{vfig2}.

Normally, in a Hamiltonian mechanical system the motion from an unstable
saddle point occurs along the corresponding unstable separatrix. However, it
is well known \cite{peyrard,OldReview,michael}\ that a spatially-periodic
setting may give rise to resonant effects, which induce an effective
dissipation, due to emission of radiation waves from a nonlinear-wave state
(vortex, in the present case). This effective dissipation, as is known \cite
{jackson,jac35,jac36,hess}, causes the vortex to spiral outward (making the
saddle point look like an unstable spiral), which is what we indeed observe
in the simulations, see Fig.\ \ref{vfig2}. Note that in other cases where
the Hamiltonian is not positive definite (such as e.g., Korteweg-de Vries --
type models), radiative losses may destabilize an equilibrium position which
is otherwise stable \cite{KdV}. Finally, in the case of $\phi =\pi /4$, the
center of the vortex is not originally at an equilibrium position of 
the potential of Eq. (\ref{veq1}), hence its drift is manifested faster, 
as is indeed seen in Fig.\ \ref{vfig3}.

\begin{figure}[tbp]
\epsfxsize=9cm \centerline{\epsffile{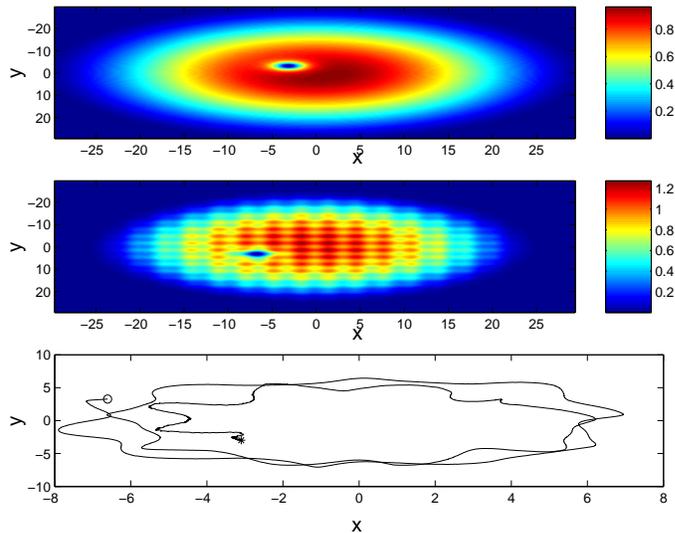}}
\caption{The top and middle panels show the vortex configuration at $t=0$
and $t=750$ ($\approx 1$ s). The bottom panel shows the trajectory of 
the vortex from $t=0$ up to $t=750$ (the two positions are marked by 
the star and circle, respectively).}
\label{vfig5}
\end{figure}

Now we turn to a different case, in which the vortex was created far from
the center of the magnetic trap (the results are displayed only for 
$\phi =0$, as it was found that the value of $\phi $ does not significantly affect
the results in this case). The motion of the vortex is displayed in Fig.\ 
\ref{vfig5}, where its outward-spiraling, due to the effective dissipation induced
by the optical lattice, is obvious. This spiral motion is combined with
jiggling induced by the potential energy landscape. Thus, we conclude that a
vortex seeded in the periphery of the BEC cloud will slowly drift towards
the edge of the cloud, where it will eventually decay into other excitations 
\cite{jac36}.

\begin{figure}[tbp]
\epsfxsize=9cm \centerline{\epsffile{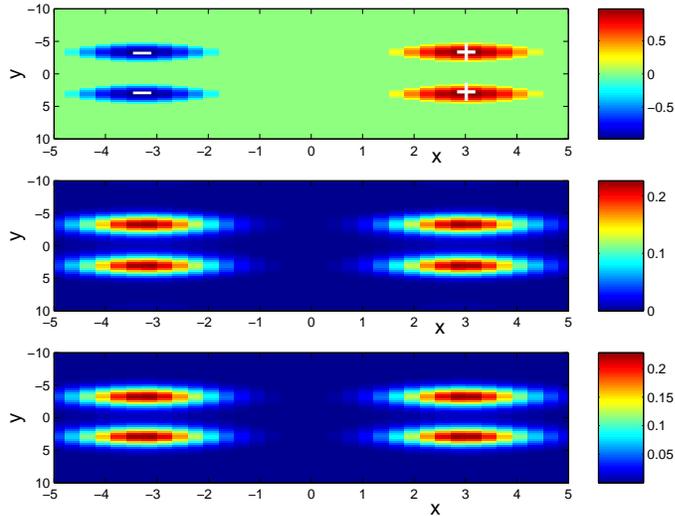}} 
\caption{The top panel shows the contour plot of the real part of the
initial configuration consisting of two pairs of pulses which are 
$\pi$ out of phase (hence the total phase change, as one goes around the
configuration is $2\protect\pi $). The pluses indicate (centers of) 
the ``up-pulses'',
whereas the minus signs the ``down-pulses'' of the initial configuration.
The middle and bottom panel show the
contour plot of the square modulus of the wave function at $t=100$ and 
$t=200 $ respectively. In the latter, we have verified that the total phase
shift around the pulse quartet is $2 \protect\pi$, hence the vortex retains
its topological charge. In this case, $V_0=2$, $k=0.5$, $\protect\phi=0$,
and $\Omega=0$.}
\label{vfig6}
\end{figure}

Finally, we present a novel type of  vortex, which is quite different
from the one considered above. Motivated by recent investigations of
vortices on 2D discrete lattices \cite{vortex1,vortex2}, we initialized a
real field configuration which is shown in the top panel of Fig.\ 
\ref{vfig6}. The configuration consists of two up-down 
(dipolar) pulse pairs, each
featuring a phase shift of $\pi $, hence a total phase shift along a contour
surrounding these pulses is $2\pi $, which corresponds to unit vorticity. In
discrete lattices, numerically exact stable stationary solutions of this
type exist \cite{vortex1}. In spatially uniform continuum models, such
stationary solutions cannot exist, but Fig.\ \ref{vfig6} shows that they may
be sustained by the OL. After shedding some radiation, the configuration
reaches a nearly steady state, although residual oscillations are observed.
Notice, once again,
 that waves radiated away due to the oscillations are absorbed by the
dissipative boundaries of the integration domain (which were imposed to
avoid artificial reflection). This dynamically stable, lattice-motivated
configuration represents, to the best of our knowledge, a novel vortex-like
structure which is particular to the system equipped with OL: without the
lattice potential, the two $\pi$-out-of-phase sets of the in-phase pulses
would separate due the pulse-pulse interactions (which are repulsive in 
the case of opposite ``parity'' pulses and attractive in the case of
same parity ones; see e.g., \cite{kkm}). However, 
the effective potential exerts a ``local force'' on the pulses, making it
possible to trap them together in a stable configuration of this type.

In conclusion, we have studied basic properties of vortices in the 2D
Gross-Pitaevskii equation with the optical lattice (OL) and magnetic trap.
The results crucially depend on the phase of the OL relative to the 
parabolic 
magnetic trap. Depending on the phase, it is possible to trap the vortex at
the center of the trap, or, on the contrary, to expel it, which is readily
explained in terms of an effective potential for the vortex
derived by means of the variational approximation. In the latter case, the
vortex moves along an unwinding spiral, which is explained by the effective
dissipation due to emission of small-amplitude waves. We also considered the
case in which the vortex was initially created far from the center of the
trap, for which a precessing motion with a slow outward-spiraling drift was
observed and qualitatively explained. Finally, a new species of a stable
vortex, specific to the system equipped with the optical lattice, was found
in the form of a bound state of two $\pi$-out-of-phase soliton pairs. The
existence of this vortex is also explained by the OL-induced potential.

Naturally, it would be of interest to further examine structures which are
inspired by features particular to periodically modulated models, and to
observe how they appear and disappear with variation of the
OL strength. Such studies are currently in progress.

This work was supported by a UMass FRG and NSF-DMS-0204585 (PGK), the
Special Research Account of the University of Athens (GT, DJF), the
Binational (US-Israel) Science Foundation, under grant No.\ 1999459 (BAM),
and San Diego State University Foundation (RCG).


\end{document}